\documentclass[pra,twocolumn,showpacs,floatfix]{revtex4}
\usepackage{amssymb}
\usepackage{graphicx}

\begin{document}

\title{Soliton stability and collapse in the discrete nonpolynomial Schr\"{o}%
dinger equation with dipole-dipole interactions}
\author{Goran Gligori\'c$^1$, Aleksandra Maluckov$^2$, Ljup\v co Had\v zievski$^1$,
and Boris A. Malomed$^3$}
\affiliation{$^1$ Vin\v ca Institute of Nuclear Sciences, P.O. Box 522,11001 Belgrade,
Serbia \\
$^2$ Faculty of Sciences and Mathematics, University of Ni\v s, P.O. Box
224, 18001 Ni\v s, Serbia \\
$^3$ Department of Physical Electronics, School of Electrical Engineering,
Faculty of Engineering, Tel Aviv University, Tel Aviv 69978, Israel}

\begin{abstract}
The stability and collapse of fundamental unstaggered bright solitons in the
discrete Schr\"{o}dinger equation with the nonpolynomial on-site
nonlinearity, which models a nearly one-dimensional Bose-Einstein condensate
trapped in a deep optical lattice, are studied in the presence of the
long-range dipole-dipole (DD) interactions. The cases of both attractive and
repulsive contact and DD interaction are considered. The results are
summarized in the form of stability/collapse diagrams in the parametric
space of the model, which demonstrate that the the attractive DD
interactions stabilize the solitons and help to prevent the collapse.
Mobility of the discrete solitons is briefly considered too.
\end{abstract}

\maketitle

\section{Introduction}

It is well established that the mean-field description of Bose-Einstein
condensates (BECs) trapped in a deep optical lattice can be reduced,
starting from the one-dimensional (1D) Gross-Pitaevskii equation with the
cubic nonlinearity, to the discrete nonlinear Schr\"{o}dinger (DNLS)
equation \cite{DNLS-BEC,DNLS-BEC-review}. However, while the DNLS equation
with the cubic on-site nonlinearity readily predicts solitons \cite{Panos},
it cannot describe the dynamical collapse, which is often observed
experimentally in the self-attractive BEC \cite{Strecker02}, \cite{Cornish}.

In the general case, the reduction of the 3D Gross-Pitaevskii equation for
the BEC trapped in a \textquotedblleft cigar-shaped" configuration to the 1D
form leads, instead of the \textquotedblleft naive" cubic Schr\"{o}dinger
equation, to ones with a nonpolynomial nonlinearity \cite{sala1,Canary}.
Such models are useful in various settings, making it possible to obtain
results in a relatively simple form, which compare well to direct
simulations of the underlying 3D equations \cite{various}. In particular,
the version of this model corresponding to the combination of the tight
transverse trap and strong optical-lattice potential acting in the
longitudinal direction takes the form of the discrete nonpolynomial Schr\"{o}%
dinger equation (DNPSE), which was introduced recently \cite{luca}. An
essential property of both the continual nonpolynomial equation and its
discrete counterpart is that they make it possible to model the collapse.

A new variety of the BEC dynamics, dominated by long-range (nonlocal)
interactions, occurs in dipolar condensates, which can be composed of
magnetically polarized $^{52}$Cr atoms, as demonstrated in a series of
experimental works \cite{Cr}. In particular, the dipole-dipole (DD)
attraction in the condensate may give rise to a specific $d$-wave mode of
the collapse \cite{d-collapse}. On the other hand, the $^{52}$Cr condensate
can be efficiently stabilized against the collapse, adjusting the scattering
length of the contact interaction by means of the Feshbach-resonance (FR)
technique \cite{experim1}. Another theoretically analyzed possibility is to
create a condensate dominated by DD interactions between electric dipole
moments induced in atoms by a strong external dc electric field \cite{dc}. A
similar situation may be realized in BEC formed by dipolar molecules \cite%
{hetmol}. In particular, recent experimental work \cite{LiCs} has reported
the creation of LiCs dipolar molecules in a mixed ultracold gas.

A possibility of making 2D solitons in dipolar condensates was predicted in
several works. In particular, isotropic solitons \cite{Pedri05} and vortices
\cite{Ami2} may exist if the sign of the DD interaction is inverted by means
of rapid rotation of the dipoles \cite{reversal}. On the other hand, stable
anisotropic solitons can be supported by the ordinary DD interaction, if the
dipoles are polarized in the 2D plane \cite{Ami1}. Solitons supported by
nonlocal interactions were also predicted and realized in optics, making use
of the thermal nonlinearity \cite{Krolik}.

A natural extension of the consideration of the dipolar BEC includes a
strong optical lattice potential, which leads to the discrete model with the
long-range DD interactions between lattice sites \cite{gpe,Santos}. In
particular, 1D unstaggered solitons in this model with the cubic on-site
nonlinearity were studied in Ref. \cite{gpe}. It was shown that the DD
interactions might enhance the solitons' stability. This conclusion suggests
a possibility of suppressing the collapse by means of the long-range DD
forces, but the study of the collapse is not possible in the discrete system
with the cubic nonlinearity. The objective of the present work is to
introduce the DD interaction into DNPSE model and analyze the effects of the
long-range interactions between dipoles localized at site of the discrete
lattice on the soliton's collapse in the sufficiently dense self-attractive
condensate trapped in the deep optical lattice potential. We here focus on
unstaggered solitons, leaving the consideration of staggered ones for
another work.

The paper is structured as follows. The discrete model including the on-site
nonpolynomial nonlinearity and long-range DD interactions is formulated in
Section II. Results for on-site and inter-site \ solitons, including the
study of their stability, and the possibility of the collapse onset, are
presented in Section III. The mobility of discrete solitons in the DNPSE
model is also briefly considered in Section III. The paper is concluded by
Section IV.

\section{The model}

Adding the long-range DD interaction to the scaled nonpolynomial Schr\"{o}%
dinger equation (in its continual form \cite{sala1,various}) leads to the
following 1D equation:

\begin{eqnarray}
i\frac{\partial F}{\partial t} &=&\left[ -\frac{1}{2}\frac{\partial ^{2}}{%
\partial z^{2}}-V_{0}\cos {(2qz)}+\frac{1-(3/2)\aleph |F|^{2}}{\sqrt{%
1-\aleph |F|^{2}}}\right] F  \nonumber \\
&+&GF(z)\int_{-\infty }^{+\infty }\frac{|F(z^{\prime })|^{2}}{|z-z^{\prime
}|^{3}}dz^{\prime }.  \label{on}
\end{eqnarray}%
Here, $V_{0}$ and $\pi /q$ are the strength and period of the longitudinal
optical lattice potential, $F(z,t)\equiv \sqrt{|\gamma |}f(z,t)$, where $f$
is the 1D mean-field wave function subject to normalization $\int_{-\infty
}^{+\infty }\left\vert f(z,t)\right\vert ^{2}dz=1$, and $\gamma =-2Na_{s}%
\sqrt{m\omega _{\bot }/\hbar }$ is the effective strength of the local
interaction, with $N $ the total number of atoms in the condensate, $a_{s}$
the scattering length of atomic collisions ($a_{s}<0$ corresponds to the
attraction), $m$ the atom mass, and $\omega _{\bot }$ the transverse
trapping frequency \cite{sala1}. Further, $\aleph \equiv\mathrm{sgn}\left(
a_{s}\right) $ is the sign of the local interaction, and $G=g\left( 1-3\cos
^{2}\theta \right) $ is the coefficient which defines the ratio of the DD
and contact interactions, where $g$ is a positive coefficient and $\theta $
the angle between the $z$ axis and the orientation of the dipoles. One
obvious case of interest in the 1D geometry corresponds to the dipoles
polarized along the $z$ axis, i.e., $\theta =0$ and $G=-2g$, when the
long-range interaction is \emph{attractive}. Another interesting case
corresponds to the orientation of the dipoles perpendicular to the $z$ axis (%
$\theta =\pi /2$ , i.e. $G=g$), which implies the repulsive DD interaction.

Assuming a sufficiently deep optical lattice potential, and approximating
the wave function by a superposition of localized Wannier modes, similar to
how it was done in Refs. \cite{luca} and \cite{gpe}, one can derive the
discrete version of Eq. (\ref{on}), i.e., the DNPSE with the DD term:

\begin{eqnarray}
i\frac{\partial F_{n}}{\partial t} &=&-C\left( F_{n+1}+F_{n-1}-2F_{n}\right)
+\frac{1-\left( 3/2\right) \aleph \left\vert F_{n}\right\vert ^{2}}{\sqrt{%
1-\aleph \left\vert F_{n}\right\vert ^{2}}}F_{n}  \nonumber \\
&-&\Gamma \sum_{n^{\prime }\neq n}\frac{\left\vert F_{n^{\prime
}}\right\vert ^{2}}{\left\vert n-n^{\prime }\right\vert ^{3}}F_{n},
\label{nordiscrete}
\end{eqnarray}%
where the linear-coupling $C$ and the ratio between DD and contact
interaction coefficients $\Gamma=G/\left|\gamma\right| $, are defined as in
Ref. \cite{gpe} ($\Gamma >0$ for attractive and $\Gamma <0$ for the
repulsive DD interactions). Two dynamical quantities are conserved by Eq. (%
\ref{nordiscrete}), \textit{viz}., norm $P=\sum_{n}\left\vert
F_{n}\right\vert ^{2}$ and Hamiltonian

\begin{eqnarray}
H &=&\sum_{n}\left[ C\left\vert F_{n}-F_{n+1}\right\vert ^{2}+\sqrt{1-\aleph
\left\vert F_{n}\right\vert ^{2}}\left\vert F_{n}\right\vert ^{2}\right.
\nonumber \\
&&\left. -\Gamma \sum_{n\neq n^{\prime }}\frac{\left\vert F_{n^{\prime
}}\right\vert ^{2}\left\vert F_{n}\right\vert ^{2}}{\left\vert n-n^{\prime
}\right\vert ^{3}}\right] .  \label{conserve}
\end{eqnarray}%
Note that the \textit{staggering transformation}, $F_{n}\equiv (-1)^{n}\exp
\left( -4iCt\right) \tilde{F}_{n}$, can be used to change the sign of $C$ if
it is negative, but this transformation cannot be used to invert the signs
of nonlinearity coefficients in Eq. (\ref{nordiscrete}).

Experimentally adjustable parameters are the relative strength of the
DD/contact interactions, $\Gamma $, and the norm of stationary wave
function, $P$. The latter may be expressed in terms of the total number of
atoms in the condensate, $N$ \cite{luca,gpe}: $P\sim N\left\vert
a_{s}\right\vert \sqrt{m\omega _{\bot }/\hbar }$. In particular, $%
a_{s}\approx 5$ nm for $^{52}$Cr atoms (far from the FR); assuming the
transverse-confinement width to be $(\sqrt{m\omega _{\bot }/\hbar })^{-1}\sim 5$ $%
\mathrm{\mu }$m, we conclude that $P\sim 1$ may correspond to $\sim 1000$
atoms in the condensate. A typical value of the relative interaction
strength, if estimated as the ratio of the effective scattering lengths
corresponding to the DD and contact interactions in the $^{52}$Cr
condensate, without the use of the FR technique, is $~\left\vert \Gamma
\right\vert \simeq 0.15$ \cite{experim1}. Actually, $\Gamma $ can be made
both positive and negative, and its absolute value may be altered within
broad limits by means of the FR \cite{Cr}. In particular, $\left\vert \Gamma
\right\vert $ can be given very large values in the experimentally possible
situation when the strength of the contact interactions is almost nullified
with the help of the FR \cite{experim1}.

Stationary solutions to Eq. (\ref{nordiscrete}), with chemical potential $%
\mu $, are sought for as $F_{n}=u_{n}\exp (-i\mu t)$, with real discrete
function $u_{n}$ satisfying a stationary equation,

\begin{eqnarray}
\mu u_{n} &=&-C\left( u_{n+1}+u_{n-1}-2u_{n}\right) +\frac{1-\left(
3/2\right) \aleph u_{n}^{2}}{\sqrt{1-\aleph u_{n}^{2}}}u_{n}  \nonumber \\
&-&\Gamma u_{n}\sum_{n^{\prime }\neq n}\frac{u_{n^{\prime }}^{2}}{\left\vert
n-n^{\prime }\right\vert ^{2}}.  \label{stationary}
\end{eqnarray}%
In the case of the attractive contact interaction ($\aleph =+1$), $u_{n}^{2}$
cannot exceed the maximum value, $\left( u_{n}^{2}\right) _{\max }=1$, as
seen from Eq. (\ref{stationary}). In fact, the presence of the singularity
in Eq. (\ref{nordiscrete}) at $\left\vert F_{n}\right\vert ^{2}=1$ makes it
possible to study the onset of collapse in the framework of this equation
\cite{luca}.

Stationary equation (\ref{stationary}) was numerically solved by an
algorithm based on the modified Powell minimization method \cite{luca,gpe}.
Initial \textit{ans\"{a}tze} used to construct on-site and
inter-site-centered discrete solitons were taken, respectively, as $\left\{
u_{n}^{(0)}\right\} =(...,\,0,\,A,\,0,\,...)$ and $(...,\,0,\,A,\,A,\,0,%
\,...)$, where $A$ is a real constant obtained from Eq. (\ref{stationary})
in the corresponding approximation. Results reported below were obtained in
the lattice composed of $101$ or $100$ sites, for the on-site and inter-site
configurations, respectively. It was checked that the results do not alter
if a larger lattice had been used.

\section{Results and discussion}

\subsection{ The case of the attractive contact interaction}

Families of fundamental unstaggered solitons of on-site and inter-site types
for the local attraction ($\aleph =+1$) and either sign of the DD
interaction were obtained from the numerical solution of Eq. (\ref%
{stationary}). In contrast to the 1D discrete model with the cubic on-site
nonlinearity, where bright unstaggered solitons exist only for sufficiently
weak repulsive DD interaction \cite{gpe}, in the present model solitons can
be also found if the repulsive DD interaction is strong. The solitons were
categorized as stable ones if they met two conditions: the slope (\textit{%
Vakhitov-Kolokolov}) criterion, according to which the slope of the $P(\mu )$
dependence must be negative (or may be very close to zero, see below), $%
dP/d\mu \leq 0$, and, simultaneously, the spectral condition, according to
which the corresponding eigenvalues, found from linearized equations for
small perturbations, must not have a positive real part \cite{gpe, stability}%
.

In order to draw general conclusions about the influence of the DD
interaction on the solitons, the analysis is presented below for two
different values of the lattice coupling constant $C$, \textit{viz}., $C=0.8$
and $C=0.2$. These cases correspond, respectively, to the proximity to the
continuum limit, and to a strongly discrete system.

\subsubsection{The quasi-continual model ($C=0.8$)}

\begin{figure}[tbp]
\includegraphics [width=7.7cm]{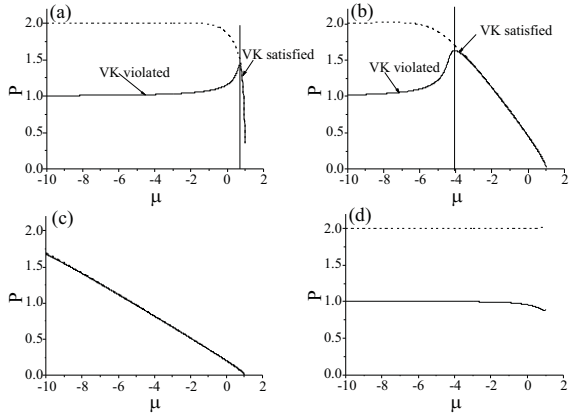}
\caption{The $P(\protect\mu )$ (norm versus chemical potential) dependencies
for families of on- and inter-site unstaggered solitons (the solid and
dashed lines, respectively) in the case of the attractive contact
interaction, for $C=0.8$ and relative strength of the DD interaction $\Gamma
=0$ (a), $5$ (b), $12$ (c), and $-5$ (d).}
\label{fig1}
\end{figure}

A global characteristic of soliton families is the $P(\mu )$ dependence,
i.e., the scaled norm as a function of the chemical potential. For $C=0.8$
and four different values of DD parameter $\Gamma $, the $P(\mu )$ curves
are displayed in Fig. \ref{fig1}. In the absence of the DD interaction, $%
\Gamma =0$, two subfamilies of on-site solitons are found, one (which
occupies a narrow interval of $\mu $) obeying the slope criterion, and the
other one violating it, in a broad interval. With the increase of the
strength of the attractive DD interactions ($\Gamma =5$), the region where
the slope condition is met spreads out, and, when the DD interaction is
dominant ($\Gamma =12$), the condition is satisfied for all on-site
solitons. On the other hand, the slope condition is fulfilled for all
inter-site solitons, regardless of the value of $\Gamma $. It is worthy to
note that the difference between the $P(\mu )$ curves for the on-site and
inter-site solitons vanishes as the attractive DD interaction strengthens.

In the case of the repulsive DD interaction, the $P(\mu )$ curves for
on-site and inter-site solitons are completely separated, see Fig. \ref{fig1}%
(d). Actually, the slope of $P(\mu )$ curves for both on-site and inter-site
solitons tends to be very small in this case, which makes the application of
the slope criterion doubtful. In these areas, the solitons feature the
amplitude close to the upper limit admitted by the model, $u_{\max }^{2}=1$,
and a very small width, corresponding to a situation when nearly all atoms
are collected in a single well of the underlying potential lattice.

The spectral stability was examined by linearizing Eq. (\ref{nordiscrete})
for small perturbations $\delta F_{n}$ around the soliton, and finding the
respective eigenvalues in a numerical form. The results of the stability
analysis for the on-site solitons, with $C=0.8$, are summarized in Fig. \ref%
{fig2}, in the form of the stability diagram in the plane of $\left( \mu
,\Gamma \right) $, where contours of constant norm $P$ for the on-site
solitons are included too. The collapse condition, $u_{\max }^{2}=1$, is
attained at the black solid line, which is, simultaneously, a stability
border. In direct simulations, the on-site solitons which are predicted to
be stable survived as long as the simulations ran [Fig. \ref{fig3}(a)],
while the solitons classified as unstable ones suffered the collapse
(destruction of the solution after attaining the level of $u_{\max }^{2}=1$)
in a finite time.

\begin{figure}[tbp]
\includegraphics [width=7.7cm]{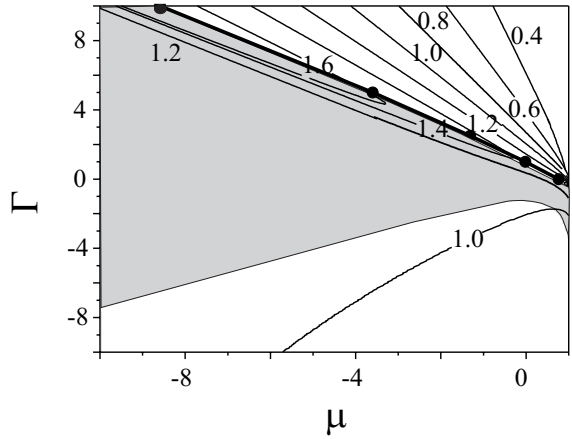}
\caption{The stability/collapse diagram for on-site solitons, as determined
by the stability analysis performed for $C=0.8$. The gray and white areas
are occupied by unstable and stable solitons, respectively. The soliton norm
keeps constant values along thin contour lines, as indicated in the figure.
The chain of bold dots connected by the black line shows a curve along which
the collapse is attained.}
\label{fig2}
\end{figure}

\begin{figure}[tbp]
\includegraphics [width=6cm]{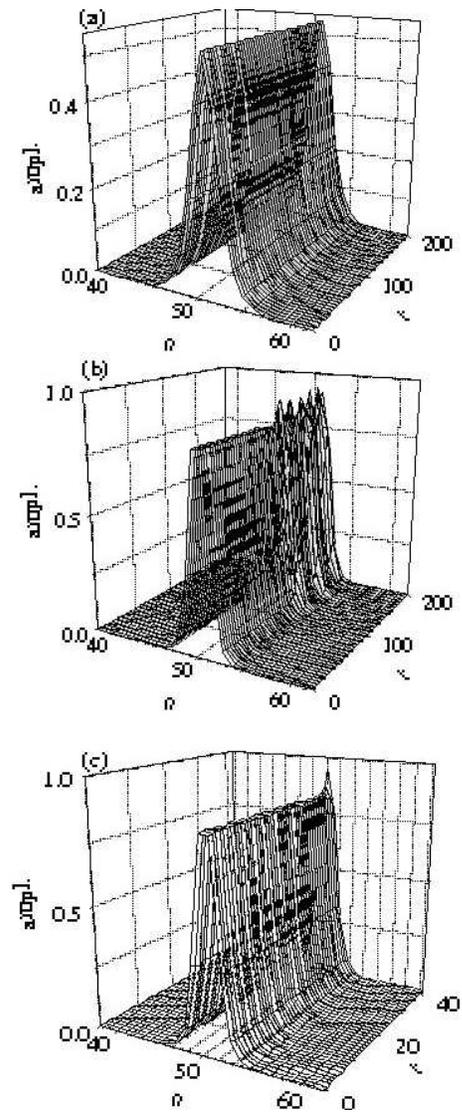}
\caption{(a) An example of a stable on-site soliton. (b,c) Development of
the instability of inter-site solitons: (b) the case when a stable on-site
soliton exists whose norm is equal to that of the unstable inter-site
sooliton. In this situation, the unstable soliton evolves into a breather
oscillating around the stable on-site soliton. (c) The case without the
stable on-site counterpart with the equal norm. In the latter case, the
unstable soliton collapses.}
\label{fig3}
\end{figure}

On the other hand, all inter-site solitons are unstable, as the spectrum of
eigenvalues for small perturbations around them always contains real
eigenvalues. However, unstable inter-site solitons which have stable on-site
counterparts with the same norm avoid the collapse, evolving into robust
breathers oscillating around the corresponding stable on-site solitons, as
shown in Fig. \ref{fig3}(b). Unstable inter-site solitons do collapse if no
stable on-site soliton with the same norm can be found, see Fig. \ref{fig3}%
(c). Therefore, the border line for the collapse of the inter-site solitons
coincides with the stability border for on-site solitons in Fig. \ref{fig2}.

For the repulsive DD interaction, the spectral stability condition for all
inter-site solitons does not hold either. Because, in this case, curves $%
P(\mu )$ for the inter-site and on-site solitons are strongly separated [Fig.%
\ref{fig1}(d)], unstable inter-site solitons do not find stable on-site
counterparts with the same norm, therefore they suffer the collapse. As for
the on-site solitons, there exists a region where the spectral stability
condition holds for them. This region expands as the repulsive DD
interaction gets stronger, although the respective curve $P(\mu )$ becomes
almost horizontal, and the slope condition cannot be accurately verified.
Direct numerical simulations confirm the predictions of the stability
analysis for the on-site solitons in this case too.

\subsubsection{The strongly discrete model ($C=0.2$)}

\begin{figure}[tbp]
\includegraphics [width=7.7cm]{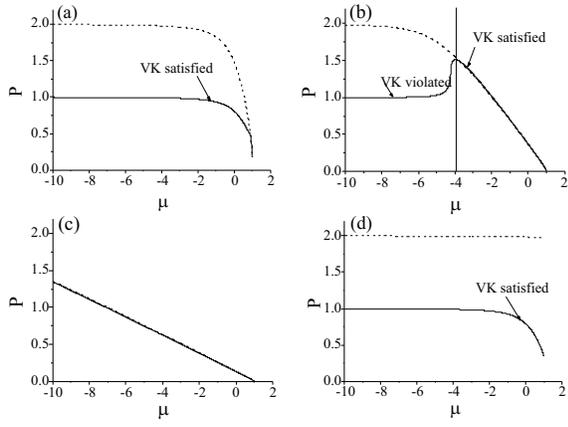}
\caption{$P(\protect\mu )$ dependencies for families of on- and inter-site
(solid and dashed lines, respectively) unstaggered solitons in the strongly
discrete model ($C=0.2$) with the attractive contact interaction. The
relative strength of the DD interaction is $\Gamma =0$ (a), $5$ (b), $15$
(c), and $-5$ (d).}
\label{fig4}
\end{figure}

In this case, the slope condition is satisfied for all on-site solitons in
the absence of the DD interaction ($\Gamma =0$), see Fig. \ref{fig4}(a). As
the strength of the attractive DD interaction grows, a pair of subfamilies
emerge, the slope-stable and unstable ones, the respective $P(\mu )$ curves
being similar to those observed in the case of the strong coupling, $C=0.8$
[Fig. \ref{fig4}(b)]. Eventually, when the attractive DD interaction becomes
dominant, the region where the slope condition is satisfied spreads over the
entire parameter space, as seen in Fig. \ref{fig4}(c).

In the model with $C=0.2$, all inter-site solitons again satisfy the slope
condition. With the strengthening of the DD interaction, the separation
between the $P(\mu )$ curves for the on-site branch, which satisfies the
slope condition, and its inter-site counterpart vanishes. On the other hand,
in the case of the repulsive DD interaction, the corresponding $P(\mu )$
curves for on-site and inter-site soliton families are strongly separated,
see Fig. \ref{fig4}(d).

\begin{figure}[tbp]
\includegraphics [width=7.7cm]{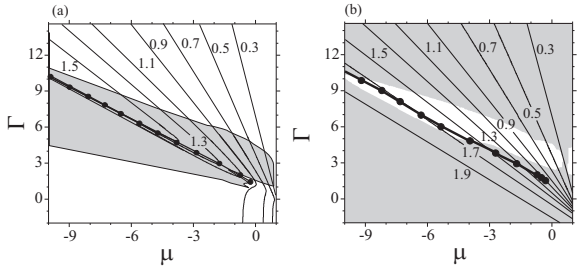}
\caption{The stability/collapse diagram for (a) on-site (a) and inter-site
(b) unstaggered solitons in the strongly discrete version of the model, with
$C=0.2$. The notation has the same meaning as in Fig. \protect\ref{fig2}.}
\label{fig5}
\end{figure}

Results of the stability analysis for the on-site solitons are summarized in
the stability diagram displayed in Fig. \ref{fig5}(a) in parametric space $%
(\Gamma ,\mu )$. Unlike the case of $C=0.8$, cf. Fig. \ref{fig2}, in the
present case there appears a region where the spectral stability condition
is violated for on-site solitons as the DD interaction grows stronger, as
well as a region where the stability condition holds for inter-site
solitons. These regions disappear again with the further growth of $\Gamma $%
. Also in contrast to the case of $C=0.8$, the line at which the collapse is
attained in the family of on-site solitons \emph{does not} coincide with the
border between stable and unstable parts of the family. Rather, the collapse
line passes trough the unstable region where the spectral-stability
condition does not hold, see Fig. \ref{fig5}(a). Direct simulations
demonstrate that unstable on-site solitons which have stable inter-site
counterparts with the equal norm evolve into breathers oscillating around
the stable counterparts. On the other hand, if unstable on-site solitons
cannot find stable inter-site counterparts with the same (or close) value of
the norm, they undergo the collapse. In fact, the existence of the two
different scenarios of the instability development -- the formation of the
breather and collapse -- explains the above-mentioned fact that the collapse
line does not coincide with the instability border. In the present case too,
direct simulations corroborate the stability of those solitons which do not
have unstable eigenvalues.

Extending the above-mentioned trend, unstable inter-site solitons which have
stable on-site counterparts with the same norm evolve into persistent
breathers, while those unstable solitons that are devoid of stable
equal-norm counterparts suffer the collapse. Accordingly, the line at which
the collapse is attained does not coincide with the instability border, as
seen in Fig. \ref{fig5}(b). It is worthy to note the existence of a
stability region for inter-site solitons in Fig. \ref{fig5}(b) which is
adjacent to the collapse line. In the latter case, the stable inter-site
solitons do not have on-site counterparts with the same value of the norm.

For the repulsive DD interactions, the spectral stability condition always
holds for on-site solitons and does not hold for inter-site modes, similar
to the case of $C=0.8$. Again, in some part of the parameter space, the
corresponding $P(\mu )$ curves for the on-site solitons are nearly
horizontal lines with zero slope, being completely separated from the $P(\mu
)$ line for the inter-site modes. Direct simulations demonstrate that the
on-site solitons are indeed stable in this case, while the unstable
inter-site solitons undergo the collapse.

\subsection{The case of repulsive contact interactions}

If the local interaction is repulsive, the existence of the unstaggered
solitons may only be supported by the attractive DD interaction. In the
model with the self-repulsive cubic on-site nonlinearity, unstaggered
solitons were found only when the relative strength of the attractive DD
interaction was large enough, $\Gamma \geq 0.4$ \cite{gpe}. In the present
model, unstaggered solitons (with large amplitudes) are obtained for smaller
values of $\Gamma $ as well. Nevertheless, general results obtained in the
present model for the case of the local repulsion are not drastically
different from those reported in the model with the cubic on-site
nonlinearity in Ref. \cite{gpe}. In particular, differences between $P(\mu )$
curves for on-site and inter-site solitons are small at all values of $\mu $%
, the slope condition is always satisfied for both types of the solitons,
and there is an exchange between regions where the spectral stability
condition is fulfilled for on-site and inter-site solitons. Further,
unstable solitons evolve into breathers oscillating around their stable
counterparts. The similarity of these results for discrete solitons in the
present DNPSE model and its cubic counterpart is not surprising, as in the
case of the local repulsion (unlike attraction) there is no dramatic
difference between the nonpolynomial and cubic nonlinearities, therefore the
competition of the local term of either type (nonpolynomial or cubic) with
the DD attraction gives rise to similar solitary modes.

\subsection{Moving discrete solitons}

Finally, we briefly summarize results concerning mobility of localized
modes, with respect to the concept of the Peierls-Nabarro barrier \cite%
{mobil}. To this end, we follow the lines of the analysis developed in Refs.
\cite{luca} and \cite{gpe}. Examination of the mobility has shown that the
Peierls-Nabarro barrier in the DNPSE model depends on the strength of the DD
interaction in the same way as it was in the discrete model with the cubic
on-site nonlinearity. Namely, in the case of the local attraction, the
repulsive or attractive DD interaction decreases or increases, respectively,
a region in plane $(P,\mu ) $ where mobile localized modes can be found. On
the other hand, in the case of the contact repulsion, all localized modes
can be set in motion by an initial kick, regardless of the value of DD
coefficient $\Gamma $. In all cases when mobile discrete solitons exist, the
vanishing of the Peierls-Nabarro barrier coincides with the disappearance of
the separation between $P(\mu )$ curves for the on-site and inter-site
soliton families, cf. Refs. \cite{luca,gpe}.

\section{Conclusion}

The purpose of this work was to achieve a better understanding on the
influence of the long-range DD (dipole-dipole) interactions on the stability
and collapse of localized nonlinear modes in the cigar-shaped Bose-Einstein
condensate trapped in a deep optical-lattice potential. To this end, we have
introduced the model based on the one-dimensional DNPSE (discrete
nonpolynomial Schr\"{o}dinger equation), which includes the contact
(on-site) and DD nonlinear terms. Both attractive and repulsive signs of the
contact and DD interactions were considered. The main conclusion is that the
presence of the attractive DD interaction enhances the soliton's stability
and helps to prevent the collapse. Our analysis was limited to unstaggered
solitons, the consideration of staggered localized modes being a subject of
a separate work.

G.G., A.M. and Lj.H. acknowledge support from the Ministry of
Science, Serbia (through project 141034).

\section*{Figures}

\end{document}